\documentclass{elsart}
                                                                                                    
\usepackage{natbib}
\usepackage{epsfig}                                                                                                    

\begin{document}
                                                                                                    
\begin{frontmatter}
                                                                                                    
                                                                                                    
\title{The lithium-rotation correlation for WTTS
          in Taurus-Auriga}
                                                                                                  
 \author[South]{L.F. Xing}\author[Oxford]{\corauthref{cor}},
 \author[South]{J.R. Shi and J.Y. Wei\thanksref{cor}}
 \corauth[cor]{Corresponding author. } 
 \address[South]{National Astronomical Observatories, Chinese Academy of Sciences, Beijing 100012, China}
 \address[Oxford]{Graduate University of the Chinese Academy of Sciences,
             Beijing 100049, China}
 \corauth[cor]{Corresponding author. }
 \ead{lfxing602@yahoo.com.cn}
 \ead{sjr@bao.ac.cn}
 \ead{wjy@bao.ac.cn}
\thanks[NSFC]{Supported by the National Science Foundation of China}
                                                                                                    
\begin{abstract}
 Surface lithium abundance and rotation velocity
 can serve as powerful and mutually complementary
 diagnostics of interior structure of stars. So far,
 the processes responsible for the lithium depletion
 during pre-main sequence evolution are still poorly
 understood. We investigate whether a correlation
 exists between equivalent widths of Li (EW(Li)) and
 rotation period (P$_{rot}$) for Weak-line  T
 Tauri stars (WTTSs). We find that rapidly rotating
 stars have lower EW(Li) and the fast burning of Li
 begins at the phase when star's P$_{rot}$ evolves
 towards 3 days among 0.9M$_\odot$ to 1.4M$_\odot$
 WTTSs in Taurus-Auriga. Our results support the
 conclusion by Piau \& Turch-Chi\'eze about a model
 for lithium depletion with age of the star and by Bouvier
 et al. in relation to rotation evolution. The turn over of
 the curve for the correlation between EW(Li) and P$_{rot}$
 is at the phase of Zero-Age Main Sequence (ZAMS).
 The EW(Li) decreases with decreasing P$_{rot}$ before
 the star reaches the ZAMS, while it decreases with
 increasing P$_{rot}$ (decreasing rotation velocity)
 for young low-mass main sequence stars. This result could
 be explained as an age effect of Li depletion and
 the rapid rotation does not inhibit Li destruction
 among low mass PMS stars.
\end{abstract} 


\end{frontmatter}

\section{Introduction}
\label{}
                                                                                                    
\appendix
 T Tauri stars (TTS) are very young ($\le 10^{8}$yr),
 low-mass (M$\le 2M_\odot$), late spectral type
 (late G to M) pre-main sequence (PMS) stars,
 still in the process of gravitational contraction. They
 are of particular interest to stellar evolution, among
 other things because they are, if sufficiently young, in
 a fully convective stage of evolution. T Tauri stars are
 usually divided into two classes, namely Classical T Tauri
 stars (CTTSs) and Weak lined T Tauri stars (WTTSs).
 The former ones are associated with a circumstellar disk, from
 which they accrete material at a rate of about 10$^{-7}$M$_{\odot}$\,yr
 $^{-1}$, while the later ones lack such disks.
                                                                                                    
 At the beginning of the T Tauri phase of the base of the convective
 zone is too cool to allow lithium burning. As the star evolves 
 on the pre-main sequence the deep regions of convective zone
 temporally exceed the $^7$Li burning point in typical stellar
 condition (~2.5$\times$10$^6$K), and surface Li can be
 transported to this region and depleted during PMS
 evolutionary phase. The $^7$Li abundancds of stars therefore offer a
 directly insight over stellar internal structure and
 evolution as it is extremely sensitive to the appearance of
 the radiative core. The surface Li abundances and surface rotation
 velocity can serve as powerful and mutually complementary
 diagnostics of interior structure\citep{Strom94}.

 Based on the result of \citet{Bouvier93} from a study of the
 connection between EW(Li) and P$_{rot}$ of 10 CTTSs and
 6 WTTSs, we concentrate on the connection between EW(Li)
 and P$_{rot}$ for more WTTSs. We did not consider
 CTTS, since accretion leads to a
 replenishment of lithium at their surface for CTTSs
 \citep{Bouvier93}, such that the Li-abundance cannot
 be used as a diagnostic for the interior. Also, the mass and the age for
 WTTSs derived from theoretical tracks and isochrones
 are more reliable than those for CTTSs. We performed a
 new and homogeneous analysis of all the Li data and
 P$_{rot}$ available in the early literature for WTTSs
 in the range of mass 0.9-1.4M$_{\odot}$.
                                                                                                    
 It is known that Li depletion increases with decreasing stellar
 mass, and with increasing metallicity \citep{Proffitt89}.
 \citet[][their Fig. 9]{Soderblom93b}
 investigated the relationship between Li abundances and
 masses for ZAMS stars in the Pleiades clusters, and found
 that, when the mass of the stars ranges between 0.9M$_\odot$ to
 1.4M$_\odot$, the Li abundances do not strongly depend on the mass.
 We selected a sample of WTTS stars in the range of mass
 0.9-1.4M$_\odot$ in Taurus-Auriga star forming regions
 (SFRs, in a same nebular, star has the similar metallicity).

 We concentrate on the connection between the EW(Li)
 and P$_{rot}$. First, since the Li I $\lambda$6707 line
 is very strong in T Tauri stars, it is on the saturated
 part of the curve of growth, so small errors in the
 measured Li I strength will lead to large error in the
 derived abundances \citep{Duncan91}. Using just the EW(Li)
 can avoid some uncertainties propagated from  uncertainties of T$_{eff}$ estimates
 which are provided by the estimates from (B-V) and (V-I)
 colors. Second, the measurements of rotation velocity
 $\upsilon \sin \it{i}$ yields only a lower limit to the
 rotation velocity due to the unknown angle of inclination,
 $\it{i}$. We use both the directly observed EW(Li) and P$_{rot}$
 instead of Li abundance and rotation velocity of stars
 in our rotation-lithium study.
                                                                                                    
 This paper is organized as follows. The periods and Li data of
 WTTSs are presented in Sect. 2.  The relationship between EW(Li)
 and P$_{\rm rot}$ and discussion provided in Sect. 3. And the
 results are summarized in Sect. 4.

\section{The Rotation periods and Li data}
\label{}
                                                                                                    
 We selected from the literature a sample of WTTSs which according to
 their position in the H-R diagram are in the mass range from
 0.9M$_\odot$ to 1.4M$_\odot$, and whose EW(Li) and
 P$_{\rm rot}$ have been determined. The location of
 the stars in the H-R diagram is shown together with
 theoretical pre-main sequence evolutionary tracks
 in Fig. 1. The theoretical pre-main sequence
 evolutionary tracks were taken from
 \citet{Cohen79}. The known binaries have been
 rejected, since the late-type binary system may be
 tidally locked rotation, which lead naturally to slower
 lithium destruction rates \citep{Maccarone05}.
                                                                                                    
\subsection{Rotation periods}
 Our sample consists of 21 WTTSs in the range of mass
 0.9-1.4M$_\odot$ in Taurus-Auriga. In table 1, the
 up-to-date rotation periods and equivalent widths
 are presented for our sample stars. The rotation
 periods of these stars are taken directly from the
 following literature:
 \begin{itemize}
 \item
 The periods of WTTSs from \citet{Bouvier93}
 are shown table 1. Our sample includes the objects
 with a multi-site photometric campaign to monitor T Tauri
 stars over more than two month by
 \citet{Bouvier93}. Two methods, namely the string-length
 method and periodogram analysis, were used to derive the
 P$_{\rm rot}$ for ten WTTS stars, most of them with
 confidence level larger than 99\%. The secondary sample
 of \citet[][their Table 4]{Bouvier93} is
 also included. Totally, we selected seven WTTSs from their
 sample.
 \item
 \citet{Xing06} selected a sample of X-ray
 sources that have been identified as WTTS stars around
 the Taurus-Auriga SFRs. They monitored the light variations
 for 22 WTTSs and obtained the P$_{\rm rot}$ for 12 stars
 using two methods: the Phase Dispersion Minimization (PDM)
 and Fourier analysis methods (PERIOD04), most of them with
 confidence level large than 99\%. We took seven WTTSs from
 their sample.
 \item
 \citet{Bouvier97} monitored the light
 variations of 58 WTTSs and derived photometric periods
 for 18 stars using 3 methods: the periodogram analysis,
 CLEAN deconvolution algorithm and string-length
 estimator. Except for RXJ0409.2+2901, all periods are
 detected at the 99\% confidence level in the periodogram.
 Five objects have been taken from their sample.
 \item
 \citet{Grankin93} presented the results of BVR
 photometry of 22 WTTSs. They obtained rotation periods
 for ten stars. Two WTTSs were selected from this sample.
 \end{itemize}
                                                                                                    
\subsection{Equivalent widths of Li}
 The equivalent widths of Li for the above stars are taken
 directly from the following literature:
 \begin{itemize}
 \item
 The EW(Li) of the \citet{Bouvier93} sample
 were taken directly from \citet{Basri91}.
 They reported the observations of strong Li I
 $\lambda$6707 line in 28 T Tauri stars in the Taurus-Auriga
 star formation complex. Line strengths were obtained using
 high resolution spectra from the Hamilten echelle at Lick
 Observatory. They have corrected the Li I equivalent
 widths for continuum veiling based on a simultaneous
 measurement of the actual veiling present.
 \item
 The EW(Li) of the \citet{Bouvier97} sample
 and three stars (RX J0430.8+2113, RX J0405.1+2632 and RX
 J0432.7+1853) of the \citet{Xing06} sample
 were taken directly from \citet{Wichmann00}.
 They presented a detailed study for Li-rich stars discovered
 by ROSAT in Taurus-Auriga SFRs, the results are based on
 high-resolution echelle spectra.
 \item
 The Li I equivalent widths of the \citet{Grankin93}
 sample and the star NTTS045251+3016 from \citet{Xing06}
 were taken from the list of \citet{Walter88},
 measured from high-dispersion spectra.
 \item
 Two stars of the \citet{Xing06} sample,
 $\lbrack LH98 \rbrack37$ and $\lbrack LH98 \rbrack53$,
 were identified as WTTSs by \citet{Li98}, and
 the Li I equivalent widths were taken from their
 results, which are based on the intermediate resolution
 spectra. The EW(Li) of HD\,287927 was taken from
 \citet{Walter86}, also based on the intermediate resolution
 spectrum.
\end{itemize}
                                                                                                    
\section{Results and discussion}\label{sec:character}
                                                                                                    
 \subsection{Li-rotation for WTTS}
 We plotted the EW(Li) versus P$_{\rm rot}$ for our sample
 stars in Fig. 2. This shows that there is a clear
 correlation between the EW(Li) and P$_{\rm rot}$, i.e. the
 rapid rotators have lower EW(Li) and the depletion of
 lithium  proceeds fast, when the rotation period of star evolves
 towards 3 days for WTTSs with mass between 0.9M$_{\odot}$
 and 1.4M$_{\odot}$ in Taurus-Auriga SFRs.
                                                                                                    
 Our results are in good agreement with the hypothesis that surface Li
 depletion takes place during PMS evolution for low-mass
 stars as a result of Li burning via (p, $\alpha$) reactions
 at low temperatures of T $\geq$ 2.6 $\times$ 10$^6$K
 \citep{King00}. These results support the conclusion by
 \citet[][their Fig. 4] {Piau02} about a model for lithium
 depletion with age of  star. Their results predict that the
 surface Li depletion will take place during PMS evolution for
 low-mass stars. They also distinguish two phases in lithium
 depletion: (1) a rapid nuclear destruction in the T Tauri
 phase before 20 Myr whatever the mass in the range between 0.8
 and 1.4M$_{\odot}$, and (2) a second phase where the
 destruction is slow and moderate, which is largely
 dependent on the hydrodynamic instability located at the
 base of the convective zone.
                                                                                                    
 It is important to know whether the EW(Li) depends on
 the masses or not. We plotted the EW(Li) against the
 masses for our sample stars in Fig. 3.  This indicates
 that the EW(Li) of our sample stars does not show any trend as
 a function of mass. It means that the EW(Li) do not
 depend on the mass of WTTSs for our sample stars.
                                                                                                    
 In order to further confirm our result, we plotted the
 EW(Li) against P$_{\rm rot}$ for WTTSs in Orion in Fig.
 4. The EW(Li) and P$_{\rm rot}$ of these stars are taken
 directly from  \citet{Alcala96} and
 \citet{Marilli05}. These selected stars
 are late G or early K-type stars whose masses are close to the
 masses of WTTS sample in Taurus-Auriga (since the mass is
 not direct observable, except in eclipsing binaries, it
 would be better to use T$_{\rm eff}$ or spectral type to
 estimate the mass  \citep{Martin97}. In the absence of a luminosity determination
 for these stars, the range of mass was estimated from
 the spectral type of the stars).  Fig. 4 shows that the correlation between
 EW(Li) and P$_{\rm rot}$ of WTTSs  (late G and early K-type)
 in Orion nebula is in a good agreement with that one of WTTSs
 (0.9M$_\odot \leq$ M $\leq$ 1.4M$_\odot$) in Taurus-Auriga.
 It is evident that the rapidly rotating stars have lower
 EW(Li), although the stars in Taurus-Auriga differ from
 that in Orion in some aspects.
                                                                                                    
\subsection{Li-rotation for young solar type stars}
                                                                                                    
 The lithium-rotation relation for ZAMS stars in Pleiades
 \citep[e.g.][]{Tschape01} and young low-mass man
 sequence stars \citep[e.g.][]{Rebolo88,Chaboyer98}
 has long been known. In \lq \lq older" clusters
 (Pleiades, Hyades) the faster rotating stars show less the
 Li depletion \citep{Tschape01}. In order to compare
 the possible effect of rotation upon lithium depletion
 between young solar-type main sequence stars and PMS,
 we take the P$_{rot}$ and EW(Li) for 7 ZAMS stars
 \citep{Krishnamurthi98, Soderblom93b, Messina01} in the
 Pleiades cluster and 12 young low-mass main sequence
 stars \citep{Rebolo88} in the Hyades cluster, and plotted
 the EW(Li) versus P$_{rot}$ of these stars with WTTSs
 in our sample in Fig. 5. The spectral types of these
 young stars were in the range between late G and early K-type
 (in the range of mass 0.9-1.4M$_\odot$, determined
 from mass-temperature relation given by
 \citet{Soderblom93c}). The EW(Li), P$_{\rm rot}$
 and other stellar properties of these stars in Pleiades
 and Hyades clusters are shown in table 3.
                                                                                                    
 Fig. 5 shows that the turn over of the lithium-rotation
 relation curve is at ZAMS phase. The EW(Li) decreases
 with decreasing P$_{rot}$ when stars young than ZAMS,
 whereas EW(Li) decrease with increasing P$_{rot}$
 (decreasing rotational velocity) when stars  are on
 the ZAMS or older.
                                                                                                    
 These results are in good agreement with the rotation
 evolution model \citep[e.g.][]{Bouvier94, Soderblom93a,
 Cemeron95, Keppens95} and Li could serve as a \lq \lq
 clock" of stellar evolution in the PMS phase \citep{Drake03}.
 The rotation evolution model and the results of these
 observations indicate that the main features of the
 rotation evolution of low-mass, late type stars are
 the strong PMS spin up from moderate rotation in the
 T Tauri phase to ultrafast rotation at ZAMS and an increase of the
 rotation period of star (spin down) with
 increasing age for main sequence stars.
                                                                                                    
 The relation between EW(Li) and P$_{rot}$ in Fig. 5 also
 shows that rapid rotators have lower EW(Li).
 This result is consistent with the lithium-age correlation
 in the sense that there is
 less lithium in the surfaces of older WTTSs (Post-TTSs)
 than in young WTTSs. e.g. the EW(Li) of older WTTSs (TAP 9
 whose location in H-R diagram very near ZAMS star) is
 lower than that of young WTTSs (TAP 57 whose
 location in H-R diagram still on Hayashi line).
                                                                                                    
\section{Summary}\label{sec:limb}
                                                                                                    
 In this work we have discussed the correlation of
 lithium-rotation. Our main conclusions can be summarized as:
 \begin{itemize}
 \item
 At least for WTTSs with mass between 0.9M$_\odot$ and
 1.4M$_\odot$ in Taurus-Auriga Nebula, there is a clear
 correlation between EW(Li) and P$_{rot}$, i.e. on average,
 rapidly rotating stars have lower equivalent widths of Li.
 That can be explained as an age effect of Li depletion
 during pre-main sequence. It is clear that rapid rotation
 does not inhibit Li depletion among low mass PMS stars. 
 \item
 The fast burnings of Li begin at the phase when the rotation period
 of the star evolves to approach 3 days. And the surface lithium
 depletion always happens during the PMS phase.
 \item
 The turn over of the lithium-rotation connection curve at the phase
 of ZAMS. The equivalent widths of Li decreases with decreasing
 rotation period when stars are younger than ZAMS, whereas the equivalent
 widths of Li decreases with increasing rotation period (decreasing
 rotation velocity) when stars evolve beyond the ZAMS.
\end{itemize}
                                                                                                    
\appendix                  

                                                                                                     \begin{table} 
\caption{Rotational periods and stellar properties for 21
         WTTSs. Reference to table 1: X\&L: \citet{Xing06} and
         \citet{Li98}; X\&W: \citet{Xing06} and \citet{Wichmann00};
         X\&Wa: \citet{Xing06} and \citet{Walter86};
         X\&G: \citet{Xing06} and \citet{Gregorio02}; B: \citet{Bouvier93};
         G\&W: \citet{Grankin96} and \citet{Walter88}; B\&W: \citet{Bouvier97} and
         \citet{Wichmann00}. The mass of stars are taken directly from above literature or
         from comparison with evolutionary tracks of
         \citet{Cohen79}. \label{tbl-1}}
\label{table:1}      
\centering                          
\begin{tabular}{l c c c c c c c r}
\hline\hline
Star & P$_{rot}$(d) & log(L$_\ast/L_\odot$) & SpT. & T$_{eff}$ & EW(Li)(m\AA) & M(M$_\odot$) &
 ref \\
\hline
&\multicolumn{6}{c}{The sample of Xing et al. photometry}& \\
\hline
$\lbrack LH98 \rbrack37$& 1.13 & 0.12 & K0IV& 5236 & 240 & 1.2  & X\&L \\
$\lbrack LH98 \rbrack53$& 0.728& -0.05& G2IV& 5792 & 230 & 1.0  & X\&L \\
  HD 287927          & 0.772& -0.12& G5  & 5554 & 200 & 0.98 & X\&G  \\
  NTTS 045251+3016   & 9.12 & 0.04 & K5  & 4034 & 580 & 0.9  & X\&Wa \\
  RX J0405.1+2632    & 1.93 & -0.34& K2  & 4897 & 219 & 0.94 & X\&W  \\
  RX J0430.8+2113    & 0.741& 0.19 & G8  & 5309 & 141 & 1.27 & X\&W  \\
  RX J0432.7+1853    & 1.55 & -0.1 & K1  & 5105 & 253 & 1.1  & X\&W \\
  \hline
&\multicolumn{6}{c}{The sample of Bouvier et al. (1993) photometry}& \\
\hline
 V1068 Tau          & 3.37 & 0.04 & K7  & 4060 & 510 & 0.9  & B \\
 V836 Tau           & 7.0  & -0.22& K7  & 4060 & 570 & 0.9  & B \\
 NTTS 045226+3013   & 2.24 & 0.17 & K0  & 5240 & 440 & 1.2  & B \\
 NTTS 034903+2431   & 1.6  & -0.3 & K5  & 4395 & 370 & 1.03 & B \\
 NTTS 041636+2743   & 5.64 & -0.04& K7  & 4060 & 600 & 0.9  & B \\
 RX J043005.1+181351& 2.7  & 0.11 & K2  & 4950 & 420 & 1.24 & B \\
 RX J043214.9+182013& 3.75 & 0.045& K7  & 4060 & 570 & 0.9  & B \\
 \hline
&\multicolumn{6}{c}{The sample of Bouvier et al. (1997) photometry}& \\
\hline
 RX J0409.2+2901    & 2.74 & 0.0  & K1  & 5105 & 413 & 1.2  &  B\&W \\
 RX J0415.4+2044    & 1.83 & 0.0  & K0  & 5236 & 270 & 1.1  &  B\&W \\
 RX J0423.7+1537    & 1.605& -0.2 & K2  & 4855 & 361 & 1.0  &  B\&W \\
 RX J0438.7+1546    & 3.07 &  0.1 & K1  & 5105 & 419 & 1.2  &  B\&W \\
 RX J0457.2+1524    & 2.39 & 0.2  & K1  & 5105 & 446 & 1.4  &  B\&W \\
\hline
&\multicolumn{6}{c}{The sample of Grankin (1993) photometry}& \\
\hline
 NTTS 041559+1716   & 2.52 & -0.4 & K7  & 4438 & 530 & 0.93 & G\&W \\
 NTTS 042835+1700   & 1.55 & -0.52& K5  & 4352 & 150 & 0.88 & G\&W \\
\hline                                   
\end{tabular}
\end{table}
%
                                                                                                    
\begin{table}
\caption{Rotation periods and stellar properties for 9
         stars in Pleiades clusters and 14
         stars in Hyades clusters. Reference to table 2:
          M\&S: \citet{Messina01} and \citet{Soderblom93b}; R:
          \citet{Rebolo88} }  
\label{table:1}      
\centering                          
\begin{tabular}{c c c c c c }        
  &\multicolumn{4}{c}{Pleiades clusters stars}&  \\
\hline\hline                 
  HII   & P$_{rot}$ &  SpT. & T$_{eff}$   & EW(Li) & ref\\
 number & (d)    &       & $^{\circ}$K & m$\AA$ &   \\
 \hline
  263  & 4.82   & G8V   & 5060 & 290 & M\&S \\
  345  & 0.84   & G8V   & 5160 & 245 & M\&S \\
  738  & 0.83   & G9V   & 5140 & 203 & M\&S \\
  1039 & 0.784  & K2V   & 4720 & 333 & M\&S \\
  2244 & 0.56   & K2.5V & 4720 & 268 & M\&S \\
  882  & 0.581  & K3V   & 4500 & 212 & M\&S \\
  1883 & 0.235  & K2V   & 4560 & 282 & M\&S \\
  3197 & 0.44   & K3v   & 4440 & 302 & M\&S \\
  1653 & 0.74   & K4.5Ve& 4220 & 108 & M\&S \\
\hline
&\multicolumn{4}{c}{Hyades clusters stars}& \\
\hline
BD+   & Period &  SpT. & T$_{eff}$   & EW(Li) & ref\\
       & (d)    &       & $^{\circ}$K & m$\AA$ &   \\
 \hline
 19 694 & 9.2  & G5  & 5460 & 32  & R \\
 17 707 & 9.1  & G8V & 5570 & 29  & R \\
 18 623 & 5.5  & G0V & 6060 & 86  & R \\
 16 589 & 6.2  & G0V & 5930 & 72  & R \\
 16 592 & 7.9  & G2V & 5840 & 69  & R \\
 18 636 & 6.1  & G5V & 5650 & 70  & R \\
 15 624 & 5.1  & G0  & 6070 & 84  & R \\
 16 601 & 8.5  & G2V & 5770 & 51  & R \\
 15 627 & 5.9  & G0  & 6200 & 85  & R \\
 16 606 & 7.4  & GV  & 5920 & 82  & R \\
 17 731 & 3.2  & G0  & 6340 & 19  & R \\
 17 734 & 11.4 & G5  & 5230 & 3   & R \\
 15 642 & 9.0  & G5  & 5540 & 15  & R \\
 15 651 & 6.5  & G0  & 5940 & 84  & R \\
\hline
\end{tabular}
\end{table}
%
\begin{figure}
\centering
\includegraphics[width=0.5\textwidth]{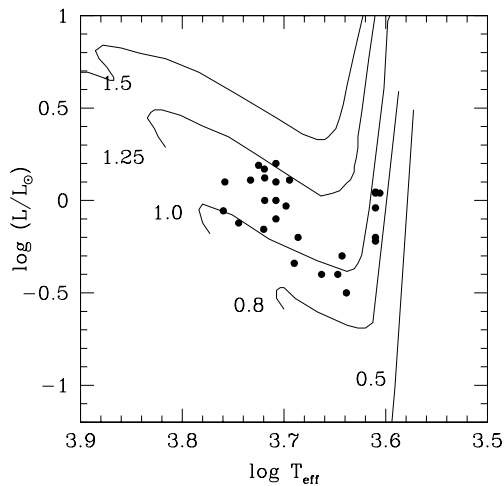}
\caption{HR diagram for our sample stars (filled circle). The pre-main-sequence
         tracks were taken from \citet{Cohen79}.
         These stars fall in the mass range between
         0.9M$_{\odot}$ to 1.4M$_{\odot}$. \label{fig2}}
\end{figure}
%
\begin{figure}
\centering
\includegraphics[width=0.5\textwidth]{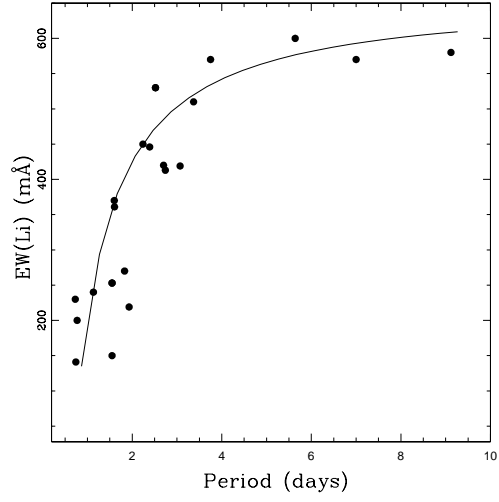}
\caption{Lithium $\lambda$6707 equivalent widths of
         WTTSs (mass range 0.9M$_\odot \leq M \leq
         1.4M_\odot$ in Taurus-Auriga SFRs) as a function
         of the rotation periods. \label{fig2}
         }
\end{figure}
%
\begin{figure}
\centering
\includegraphics[width=0.5\textwidth]{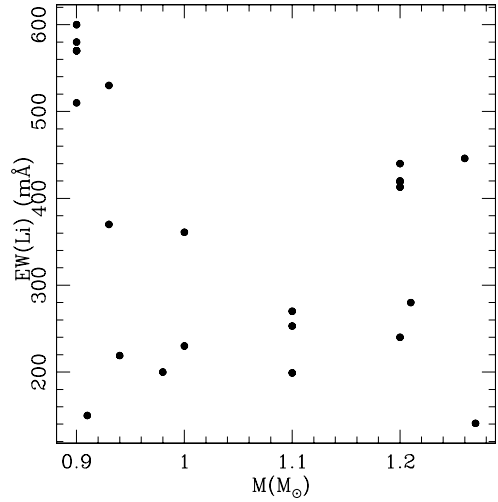}
\caption{Lithium $\lambda$6707 equivalent widths of
         WTTS (mass range cover by 0.9M$_\odot \leq M \leq
         1.4M_\odot$ in Taurus-Auriga SFRs) as a function of their masses.
          \label{fig3}
         }
\end{figure}
                                                                                                    
\begin{figure}
\centering
\includegraphics[width=0.5\textwidth]{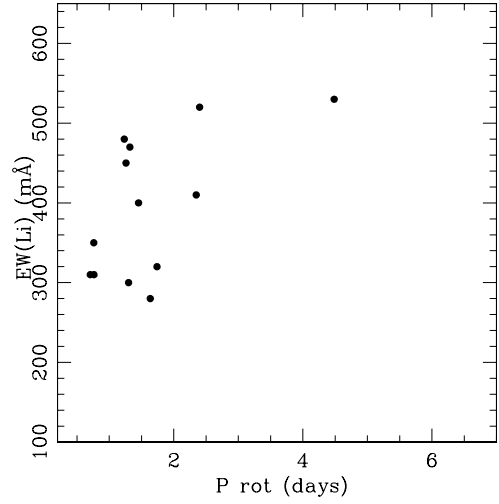}
\caption{Lithium $\lambda$6707 equivalent widths for
         WTTSs in Orion nebula as a function of their rotational periods.
         The spectral type of sample stars ranges from G9 to K5.
          \label{fig4}
        }
\end{figure}
                                                                                                    
\begin{figure}
\centering
\includegraphics[width=0.5\textwidth]{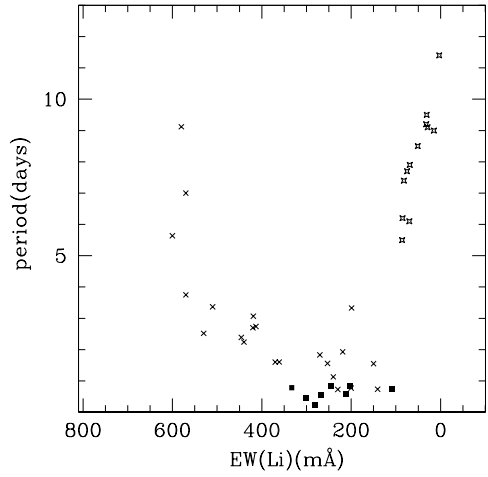}
\caption{The lithium $\lambda$6707 equivalent widths as a function of the
         rotational periods of WTTSs of mass 0.9M$_\odot \leq M \leq
         1.4M_\odot$ in Taurus-Auriga SFRs (crosses), ZAMS stars between late
         G and early K-type in Pleiades cluster (filled square) and young solar-like
         stars between late G and early K-type in Hyades cluster (stars).
         \label{fig5}
         }
\end{figure}

\end{document}